\documentclass[a4paper,aps,prl,reprint]{revtex4-1}

\usepackage{hyperref}

\usepackage{amsfonts,amssymb,amsbsy,latexsym,amsmath,tabulary,graphicx,time}
\usepackage[utf8]{inputenc}
\usepackage{multirow,morefloats,floatflt,cancel,tfrupee}

\usepackage{colortbl}
\usepackage{xcolor}
\usepackage{pifont}

\usepackage[T1]{fontenc}

\usepackage{float}

\newcommand{\fig}{Fig.~}
\newcommand{\figs}{Figs.~}

\newcommand{\eq}[1]{Eq.~(\ref{#1})}
\newcommand{\eqs}[1]{Eqs.~(\ref{#1})}
\newcommand{\jeq}[1]{(\ref{#1})}

\begin{document}

\title{Magnetic control of flexible thermoelectric devices based on macroscopic 3D interconnected nanowire networks}

\author{Flavio Abreu~Araujo}
\author{Tristan da~C\^{a}mara Santa~Clara Gomes}
\author{Luc Piraux}
\affiliation{Institute of Condensed Matter and Nanosciences, Universit\'{e} catholique de Louvain\unskip, Place Croix du Sud 1, Louvain-la-Neuve, 1348, Belgium}

\begin{abstract}
	Spin-related effects in thermoelectricity can be used to design more efficient refrigerators and offer novel promising applications for the harvesting of thermal energy. The key challenge is to design structural and compositional magnetic material systems with sufficiently high efficiency and power output for transforming thermal energy into electric energy and vice versa. Here, the fabrication of large-area 3D interconnected Co/Cu nanowire networks is demonstrated, thereby enabling the controlled Peltier cooling of macroscopic electronic components with an external magnetic field. The flexible, macroscopic devices overcome inherent limitations of nanoscale magnetic structures due to insufficient power generation capability that limits the heat management applications. From properly designed experiments, large spin-dependent Seebeck and Peltier coefficients of $-9.4$~$\mu$V/K and $-2.8$~mV at room temperature, respectively. The resulting power factor of Co/Cu nanowire networks at room temperature ($\sim7.5$~mW/K$^2$m) is larger than those of state of the art thermoelectric materials, such as BiTe alloys and the magneto-power factor ratio reaches about 100\% over a wide temperature range. Validation of magnetic control of heat flow achieved by taking advantage of the spin-dependent thermoelectric properties of flexible macroscopic nanowire networks lay the groundwork to design shapeable thermoelectric coolers exploiting the spin degree of freedom.
	\def\keywordstitle{Keywords}\\%
\end{abstract}

\maketitle
	
\section{Introduction}
	
Innovative spin-based transport mechanisms are crucial steps in developing the next generation of thermoelectric materials \cite{He2017}. With this perspective, coupling heat-driven transport with spintronics is at the heart of the rapidly emerging field of spin caloritronics \cite{Bauer2012, Boona2014}. Previous studies on nanoscale metal structures, magnetic tunnel junctions and magnetic insulators have led to the observation of various spin-enabled mechanisms that may differ significantly from conventional thermoelectrics effects such as spin Seebeck effects \cite{Uchida2010_Nature, Jaworski2010}, thermally driven spin injection \cite{Slachter2010} and thermal assisted spin-transfer torque \cite{Hatami2007, Pushp2015}. However, the low conversion efficiency of the observed effects, the weak power output of the spin caloritronic devices and the lack of reproducibility in the experiments has limited the application as heat harvesters. Although there have been some initial studies on the thermoelectric analogues of giant magneto-resistance (GMR) in magnetic multilayers with current in-plane configuration \cite{Conover1991, Piraux1992, Shi1993, Shi1996}, the effects of interfaces make the interpretation of the results more delicate than in the simpler CPP (current-perpendicular-to-plane) configuration \cite{Tsymbal2001}. Indeed, in the limit of no-spin relaxation, most of the CPP-GMR data can be understood using a simple two-current series-resistor model, in which the resistance of layers and interfaces simply add and where 'up' and 'down' charge carriers are propagating independently in two spin channels with large spin asymmetries of the electron's scattering \cite{Lee1992, Bass2016}.
	
Similarly, the spin-dependent thermoelectric effects exploit the fact that the Seebeck coefficients for spin-up and spin-down electrons are also different. The diffusion thermopower arises from the diffusion of charge carriers opposite to the temperature gradient. It is related to the energy dependent conductivity of the material $\sigma(\epsilon)$ by Mott's formula:
\begin{equation}
	S = -eL_0T \left. \left( \frac{d \log \sigma(\epsilon)}{d \epsilon}  \right) \right|_{\epsilon = \epsilon_\text{F}}\text{,}
	\label{Eq1}
\end{equation}
with $L_0 =$ 2.44 $\cdot$ 10$^{-8}$ V$^{2}$K$^{-2}$ the Lorenz number and $e$ the electron charge (positive). According to Einstein's relation for a metal or alloy with isotropic properties, the conductivity is proportional to the density of states $N(\epsilon)$ and to the scattering time $\tau(\epsilon)$, where both terms in \eq{Eq1} are to be evaluated at the Fermi level $\epsilon_\text{F}$. Because of the pronounced structure of the d-band and the high energy derivative of the density of states at the Fermi level in 3d ferromagnetic metals, large diffusion thermopowers are obtained (e.~g. $S \approx$ $-30$ $\mu$V/K in cobalt at room temperature (RT)). Moreover, significantly different Seebeck coefficients for spin-up and spin-down electrons, $S_\uparrow$ and $S_\downarrow$, are expected because the d-band is exchange-split in these ferromagnets, as suggested from previous works performed on dilute magnetic alloys \cite{Farrell1970, Cadeville1971}.
	
To date, most of the investigations of thermoelectric transport in CPP-GMR systems were performed on lithographically defined nanopillars, single nanowire (NW) and parallel NW arrays \cite{Gravier2004, Gravier2006, Flipse2012, Bohnert2014}. Spin-dependent Seebeck and Peltier effects were also recently reported in magnetic tunnel junctions \cite{Liebing2011, Walter2011}. So far, major experimental issues are the insufficient power generation capability in such magnetic nanostructures and the lack of established methods for reliable measurements of spin caloritronic material parameters. Indeed, it is difficult to determine and/or eliminate contact thermal resistance, an important error source, and simulations are often required to estimate the temperature gradient over the multilayer stacks \cite{Liebing2011, Walter2011}. Therefore, accurate determination of $S_\uparrow$ and $S_\downarrow$ still remains challenging and from these previous works, only relatively small values of $S_\uparrow - S_\downarrow$ ranging from -1.8~$\mu$V/K to -4.5~$\mu$V/K for cobalt and permalloy, respectively, were indirectly estimated from measurements performed on spin valves devices \cite{Slachter2010, Dejene2012}. In addition, observations of the magnetically controlled Peltier cooling effect in nanostructures have not been possible because of the dominant Joule heating effect compared to the Peltier effect together with the technical challenges in detecting extremely small temperature changes. So, in spite of the practical significance, all these experimental issues impose strong restrictions on applications in the burgeoning field of spin-caloritronics.
	
Very recently, we have developed an experimental platform that allows for the observation of magnetic control of heat flux in custom-fabricated spin caloritronic devices. Specifically, we fabricated macroscopic interconnected networks made of CoNi/Cu multilayer NWs by direct electrodeposition into 3D nano-porous polymer host membranes \cite{Camara-Santa-Clara-Gomes2019} (see Experimental Section). A significant advantage of our approach is the achievement of highly efficient flexible and shapeable spin caloritronics devices. Moreover, since there is no sample size limitation, the fabrication method is directly expandable into nanowire network films with much larger dimensions.
	
In this work, we report the results of spin caloritronic measurements performed on Co/Cu nanowire networks. These multilayers are easier to fabricate by potentiostatic electro-chemical deposition than CoNi/Cu. Also, cobalt exhibits the largest thermoelectric power factor (PF $= S^2\sigma \approx$ 15 mW/K$^2$m for bulk material with $\sigma$ the electrical conductivity) at RT \cite{Vandaele2017}. The power factor is the physical parameter that relates to the output power density of a thermolectric material. Moreover, the spin diffusion length, which is the mean distance that electrons diffuse between spin-flipping collisions, is significantly larger in Co than in magnetic alloys \cite{Bass2016, Dejene2012}. Therefore, restrictions on layer thicknesses are less for Co in order to observe large magneto-transport effects. Using centimetre-scale Co/Cu nanowire network films we demonstrated the magnetic field control of Peltier cooling of macroscopic electronic components and we extracted accurately key spin-dependent material parameters over a wide temperature range from thermoelectric measurements. 
	
\section{Results and Discussion}
	
\begin{figure*}[!htbp]
	\IfFileExists{Fig1.pdf}{\includegraphics[scale=1.4]{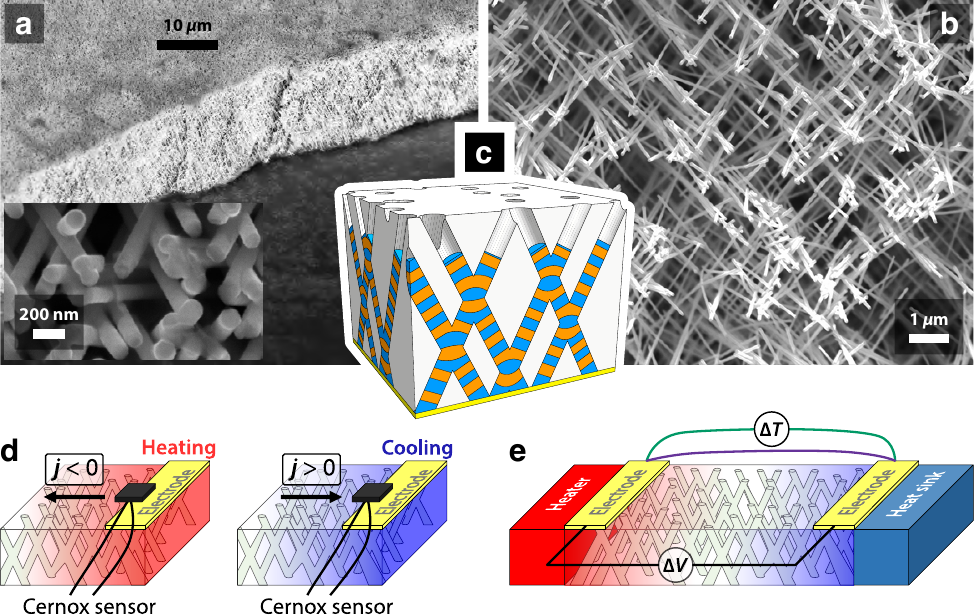}}{}
	\caption{(a-b) SEM images of self-supported interconnected nanowire (NW) network with different magnifications, diameters and packing densities. (a) Low-magnification image showing the 50$\circ$-tilted view of the macroscopic NW network film with 105 nm diameter and 22\% packing density. The inset shows the nanowire branched structure at higher magnification. (b) Low magnification image showing the top view of the NW network with 80 nm diameter and 3\% packing density. (c) Schematic of the interconnected NW network with alternating magnetic and non-magnetic layers embedded within the 3D nanoporous polycarbonate template. (d-e) Experimental set-ups for spin caloritronics measurements. (d) Device architecture and measurement configuration for the Peltier cooling/heating detection. The Cernox temperature sensor is used to probe the local temperature change at the electrode - nanowire network junctions induced by the electric current flow (see Experimental Section for details). (e) Device configuration to measure the Seebeck coefficient and the magneto-thermoelectric effect. Heat flow is generated by a resistive element and a thermoelectric voltage $\Delta V$ is created by the temperature difference $\Delta T$ between the two metallic electrodes that is measured by a thermocouple (see Experimental Section for details). Device dimensions in (d-e) are 15 mm long, 5 mm wide and 22 $\mu$m thick and the colour represents the generated temperature profile in the NW networks. The gold electrodes are 2 mm wide.}
	\label{Fig1}
\end{figure*}

The interconnected nanoporous templates have been prepared by performing a sequential two-step exposure to energetic heavy ions, at angles of +25$^\circ$ and -25$^\circ$ with respect to the normal of the polycarbonate (PC) membrane surface \cite{Araujo2015} (see Experimental Section). Two different 22 $\mu$m thick polymer membranes were used in this study with diameters 80 and 105 nm and very different porosity characteristics. The spin-caloritronic devices are networks of interconnected NWs embedded inside the 3D polymer membranes. The crossed NWs with multilayer structure of Co/Cu were grown in the host porous templates by pulsed electrochemical deposition \cite{Piraux1994}. The thickness of the bilayers was set as 15 nm with approximately the same thickness for the Co and Cu layers. \fig\ref{Fig1}(a-b) displays the morphology of self-supported interconnected NW networks obtained after template dissolution, whereas \fig\ref{Fig1}c illustrates the 3D multilayered architecture. The nanowire network films are the inverse replicas of the 3D nano-porous polymer templates. As schematically outlined in \fig\ref{Fig1}(d-e), we developed two experimental set-ups allowing to simultaneous determine the Peltier heating and cooling power generated at the NW network/metal electrode junction (d) and to measure the magneto-resistance and Seebeck coefficient (e) of the flexible NW network films (see Experimental Section).

\subsection{Spin-dependent thermoelectric parameters of Co/Cu nanowire networks}
	
\begin{figure*}[!htbp]
	\IfFileExists{Fig2.pdf}{\includegraphics[scale=0.55]{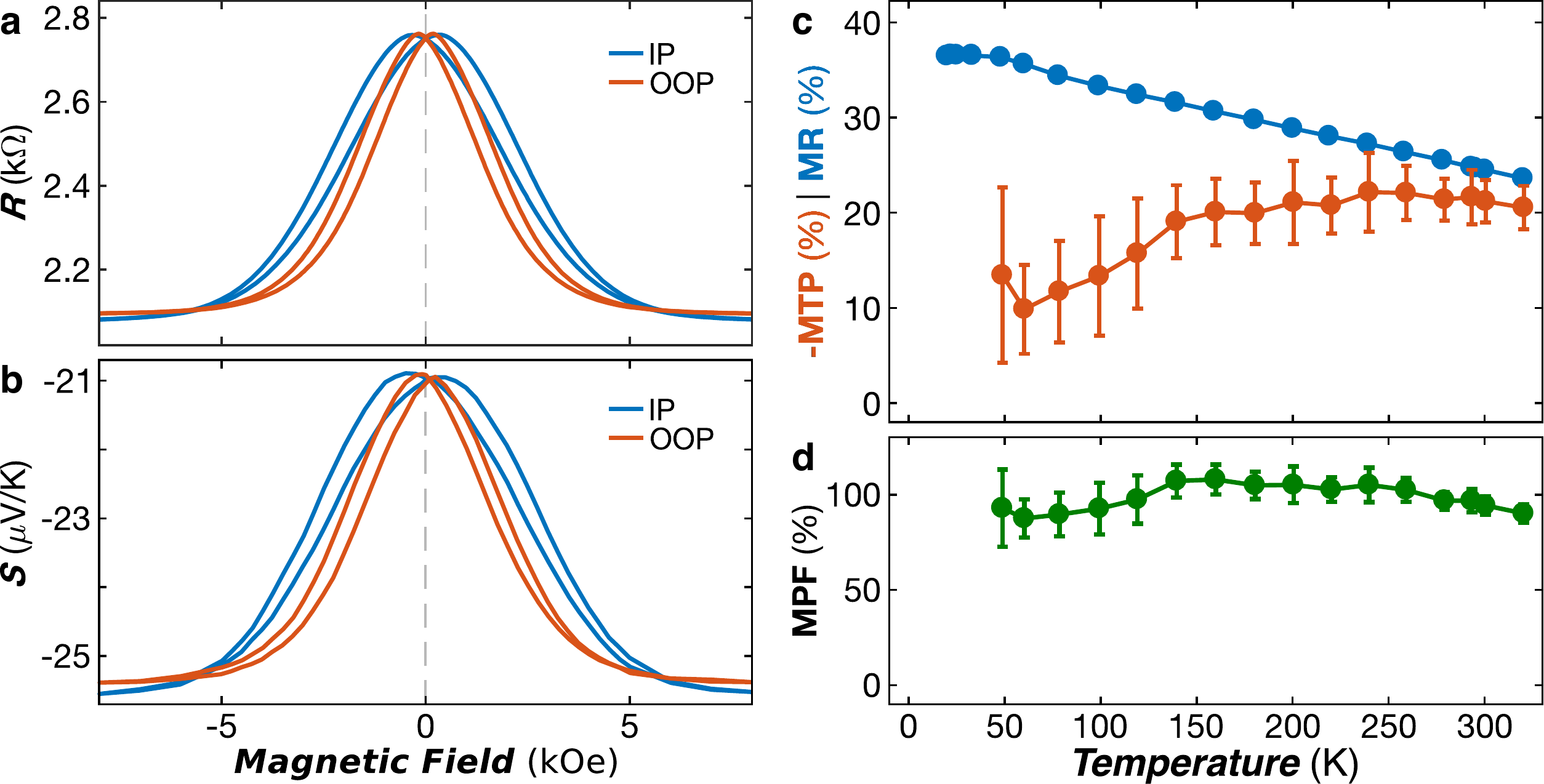}}{}
	\caption{{Measured thermoelectric characteristics. (a-b) Electrical resistance (a) and Seebeck coefficient (b) of a Co/Cu NW network 80 nm in diameter and 3\% packing density showing similar magnetic field dependence corresponding to MR $=$ 24.7\% and MTP $=$ $-$22.3\% at room temperature. The curves in (a-b) were obtained with the applied field in-plane (IP - in blue) and out-of-plane (OOP - in red) of the NW network film. (c) MR ratio and -MTP as a function of temperature with the field applied in the plane of the NW network film. (d) Magneto-power factor as a function of temperature obtained using the MR and MTP data in (c) and MPF $= (1 - \text{MTP})^2/(1 - \text{MR}) - 1$. The error bars in (c-d) reflect the uncertainty of the electrical and temperature measurements and is set to two times the standard deviation, gathering 95\% of the data variation.}}
	\label{Fig2}
\end{figure*}

As shown in \fig\ref{Fig2}(a-b), the resistance and thermopower of the low-packing density Co/Cu NW sample show the same magnetic field dependencies and similar relative changes of $\sim$25\% at $H = 8$~kOe at RT. The samples are nearly magnetically isotropic, as observed from the magneto-transport curves obtained with the applied magnetic field along the out-of-plane (OOP) and in-plane (IP) directions of the NW network films (see \fig\ref{Fig2}(a-b)). This behaviour corresponds to the one expected considering the crossed nanowire architecture and magneto-static arguments when using similar magnetic and nonmagnetic layer thicknesses[30]. In the following, only the measurements obtained in the plane of the NW network films are reported. Interestingly, the fabrication method appears as a very convenient approach for large scale production of CPP-GMR films based on crossed NW networks, exhibiting very large GMR responses (defined as $R_{AP} /R_P - 1$, with $R_{AP}$ and $R_P$ the corresponding resistances in the high- and low-resistance states, respectively) reaching values of 33\% at RT and 58\% at $T =$ 10 K in the present study. Besides, the CPP geometry of the device is suitable for spin caloritronic purposes since the measured thermopower on the Co/Cu NW network in the saturated state is not significantly different from that found in pure cobalt material, as expected from conventional rules for a stacking arrangement in series. Indeed, assuming $t_\text{Co} \approx t_\text{Cu}$, the total Seebeck coefficient of the system can be expressed as $S_{\text{Co}/\text{Cu}} \approx (S_\text{Co} \rho_\text{Co} + S_\text{Cu} \rho_\text{Cu})/(\rho_\text{Co} + \rho_\text{Cu})$, where $S_\text{Co} \rho_\text{Co} >> S_\text{Cu} \rho_\text{Cu} $.

The absolute value of the magnetothermopower MTP $= (S_\text{AP} - S_\text{P})/S_\text{AP}$, with $S_\text{AP}$ and $S_\text{P}$ the corresponding thermopowers in the high- and low-resistance states, respectively, shows a similar
increase with decreasing temperature as the MR ratio (defined as MR $= (R_\text{AP} - R_\text{P})/R_\text{AP})$ for temperatures in the range 250 - 320 K (see \fig\ref{Fig2}c). However, below T $\sim$ 250 K, the MTP exhibits
a less pronounced effect and reach a minimum value around -10\% at $T =$ 60 K. Calculations of the thermoelectric power factor at RT lead to $\text{PF}_\text{AP} = S_\text{AP}^2 /\rho_\text{AP} \approx$ 3.8~mW/K$^2$m and $\text{PF}_\text{P} = S_\text{P}^2/\rho_\text{P} \approx$ 7.5~mW/K$^2$m (See Supporting Information Section S1 for the estimation of the resistivity values), which are comparable values or even larger than the PF of the widely used thermoelectric material, bismuth telluride (in the range 1-5 mW/K$^2$m)\cite{Yamashita2002}. Besides, the PF values obtained for Co/Cu NW networks embedded in polymer membranes are at least one order of magnitude larger than those of flexible thermoelectric films based on optimized conducting polymers \cite{Bubnova2011}. The magneto-power factor (MPF $= (\text{PF}_\text{P} - \text{PF}_\text{AP})/\text{PF}_\text{AP})$ reaches $\sim$100\% at RT and shows a weak temperature variation down to $T \approx$ 50 K, as also shown in \fig\ref{Fig2}d. The efficiency of a material's thermoelectric energy conversion is determined by its figure of merit $ZT = S^2 \sigma T/ \kappa$ with $\kappa$ the thermal conductivity. In a previous work, M. Ou et al \cite{Ou2008} have measured the thermal conductivity of a suspended nickel nanowire for $T =$ 15–300 K. While the Lorenz ratio $L = \kappa/\sigma T$ departs from the Sommerfeld value ($L_0 =$ 2.45 $\cdot$ 10-8 V$^2$/K$^2$) at low temperatures, $L$ was found to be equal to $L_0$ with a 5\% margin of error above $T =$ 50 K. Due to the very low thermal conductivity of polycarbonate ($\kappa =$ 0.2 W/m$\cdot$K at RT), the contribution of the polymer matrix to heat transport is much smaller than that of the metallic nanowire network. Indeed, assuming that the Wiedemann-Franz law holds for metallic nanowires, an estimate of the RT electronic thermal conductivity gives $\kappa_\text{E} =$ 63 W/m$\cdot$K for an AP configuration and $\kappa_\text{E} =$ 84 W/m$\cdot$K for a P configuration. In this case, the figure of merit is reduced to $ZT = S^2/L_0$, thus leading to $ZT \approx$ 2.5$\cdot$ 10$^{-2}$ for the Co/Cu NW network sample at RT. Although the figure of merit is more than one order of magnitude smaller than those of state-of-the-art thermoelectric materials ($ZT \approx$ 1 in BiTe alloys), it is comparable to those of thermocouple alloys ($ZT \approx$ 6$\cdot$ 10$^{-2}$ and $ZT \approx$ 1.4$\cdot$10$^{-2}$ in constantan and chromel, respectively) and can be used in applications for devices with low energy requirements when the supply of heat essentially is free as with waste heat. Furthermore, considering the energy conversion from heat to electric power, the $ZT$ of the proposed device is much larger than that of a spin Seebeck power generator ($ZT \approx$ 10$^{-4}$) based on a device using a two-step conversion process and the inverse spin Hall effect to convert spin current to charge current in non-magnetic materials \cite{Cahaya2014}.

\begin{figure*}[!htbp]
	\IfFileExists{Fig3.pdf}{\includegraphics[scale=0.45]{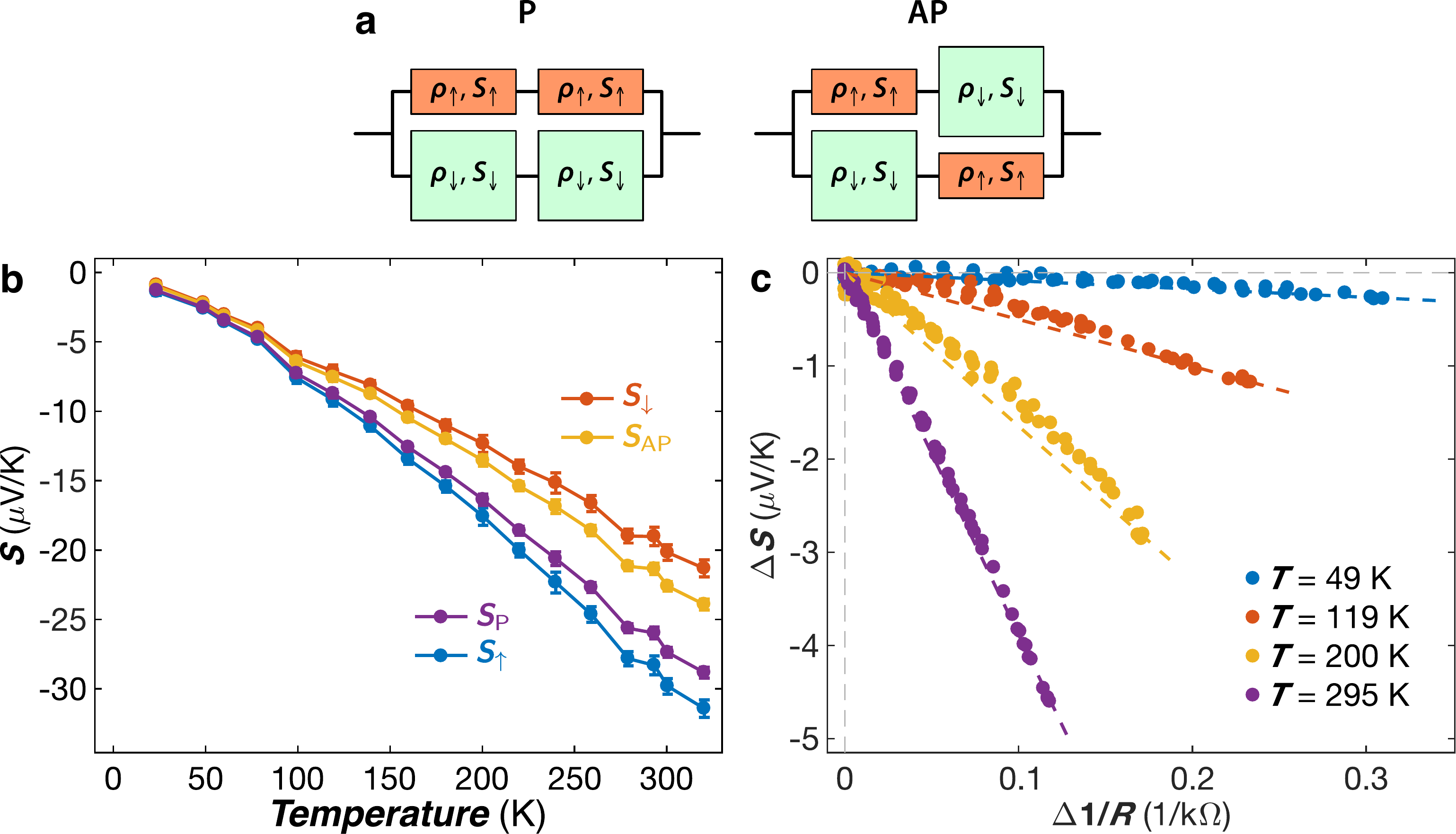}}{}
	\caption{{(a) The two-current model for the resistivity and the thermopower considering both parallel (P) and antiparallel (AP) magnetic configurations. (b) Measured Seebeck coefficients at zero applied field $S_\text{AP}$ (orange circles) and at saturating magnetic field $S_\text{P}$ (violet circles)of a Co/Cu NW network 80 nm in diameter and 3\% packing density, along with the corresponding calculated $S_\uparrow$ (blue circles) and $S_\downarrow$ (red circles) from \eqs{Eq4} and \jeq{Eq5} (see text). (c) Linear variation of $\Delta S(H)$ vs $\Delta (1/R(H))$ at different measured temperatures illustrating the Gorter-Nordheim characteristics. The dashed lines correspond to the theoretical relation shown in \eq{Eq6} (see text). The error bars in (c) reflect the uncertainty of the electrical and temperature measurements and is set to two times the standard deviation, gathering 95\% of the data variation.}}
	\label{Fig3}
\end{figure*}

Assuming that the layers of the magnetic multilayers are thin compared to the spin-diffusion lengths, we may use the simple two-current model, for which separate resistivities $\rho_\uparrow$ and $\rho_\downarrow$ and Seebeck coefficients $S_\uparrow$ and $S_\downarrow$ are defined for majority and minority spin channels \cite{Valet1993, Shi1993, Gravier2004}. Accord- ing to the usual rule when the currents split to flow along two parallel paths (See \fig\ref{Fig3}a), the corresponding thermopowers $S_\text{AP}$ and $S_\text{P}$ are simply given by \cite{Shi1993}:
\begin{equation}
S_\text{AP} =  \frac{S_\uparrow \rho_\uparrow + S_\downarrow \rho_\downarrow}{\rho_\uparrow + \rho_\downarrow} \text{,}
\label{Eq2}
\end{equation}
and:
\begin{equation}
S_\text{P} =  \frac{S_\uparrow \rho_\downarrow+ S_\downarrow \rho_\uparrow}{\rho_\uparrow + \rho_\downarrow} \text{,}
\label{Eq3}
\end{equation}
Therefore, the spin-dependent Seebeck coefficients, $S_\uparrow$ and $S_\downarrow$ can be expressed as follows:
\begin{equation}
S_\uparrow =  \frac{1}{2} \left[ S_\text{AP} \left( 1 - \beta^{-1} \right)  + S_\text{P} \left( 1 + \beta^{-1} \right) \right] \text{,}
\label{Eq4}
\end{equation}
\begin{equation}
S_\downarrow =  \frac{1}{2} \left[ S_\text{AP} \left( 1 + \beta^{-1} \right)  + S_\text{P} \left( 1 - \beta^{-1} \right) \right] \text{,}
\label{Eq5}
\end{equation}
where $\beta = (\rho_\downarrow-\rho_\uparrow)/(\rho_\downarrow+\rho_\uparrow)$ denotes the spin asymmetry coefficient for resistivity. From \eqs{Eq4} and \jeq{Eq5}, it can be easily deduced that $S_\uparrow = S_\text{P}$ and $S_\downarrow = S_\text{AP}$ in the limit of an extremely large MR ratio ($\beta \rightarrow$ 1). The temperature evolutions of $S_\text{AP}$, $S_\text{P}$, $S_\uparrow$ and $S_\downarrow$ are shown in \fig\ref{Fig3}b. The Seebeck coefficients decrease almost linearly with decreasing temperature, which is indicative of the dominance of diffusion thermopower. The estimated values at RT are: $S_\uparrow$ = $-27.9$ $\mu$V/K and $S_\downarrow = -18.5$ $\mu$V/K using $\beta = (\text{MR}^{1/2}) = 0.50$, which are similar to those previously reported in bulk Co ($S_\uparrow = -30$ $\mu$V/K and $S_\downarrow = -12$ $\mu$V/K)\cite{Cadeville1971}. The value of $\beta$ so obtained is also in agreement with previous estimates from the CPP-GMR experiments performed on Co/Cu multilayers \cite{Bass2016}. In contrast, our value for $S_\uparrow - S_\downarrow$ of $-$9.4 $\mu$V/K is much larger than the one of $-1.8$ $\mu$V/K extracted from measurements performed on a Co/Cu/Co nanopillar spin valve using a 3D finite-element model \cite{Slachter2010, Dejene2012}. The value of Sup-S down obtained for Co/Cu nanowire networks is only slightly smaller than the one recently reported for CoNi/Cu system ($-11.5$ $\mu$V/K at RT)\cite{Camara-Santa-Clara-Gomes2019}.

Further evidence that the thermopower is dominated by electron diffusion over the whole temperature range investigated is presented in \fig\ref{Fig3}c. Defining the diffusion thermopower $S(H) = e L_0 T \rho'(H)/\rho(H)$ by Mott's formula (\eq{Eq1}) with $\rho'(H) = \left(d\rho(H)/D\epsilon \right)_{\epsilon = \epsilon_\text{F}}$ the derivative of the electrical resistivity with respect to the energy, evaluated at the Fermi level, one can write $S_\text{AP} = eL_0T\rho'_\text{AP}/\rho_\text{AP}$ and $S_\text{P} = eL_0T \rho'_\text{P}/\rho_\text{P}$. Then, the following expression describing the inverse relationship between the field-dependent thermopower and electrical resistance can be easily obtained:
\begin{equation}
S(H) =  A + \frac{B}{R(H)} \text{,}
\label{Eq6}
\end{equation}
where $A = (S_\text{P}R_\text{P}-S_\text{AP}R_\text{AP})/(R_\text{P}-R_\text{AP})$ and $B = (R_\text{P} R_\text{AP}(S_\text{AP}-S_\text{P}))/(R_\text{P}-R_\text{AP})$. This expression corresponds to an equivalent form of the Gorter-Nordheim relation for diffusion thermopower in metals and alloys \cite{Blatt1976} and has been observed at different temperatures in the interconnected network made of Co/Cu NWs, as shown in \fig\ref{Fig3}c.

\subsection{Magnetic control of heat currents}

\begin{figure*}[!htbp]
	\IfFileExists{Fig4.pdf}{\includegraphics[scale=0.45]{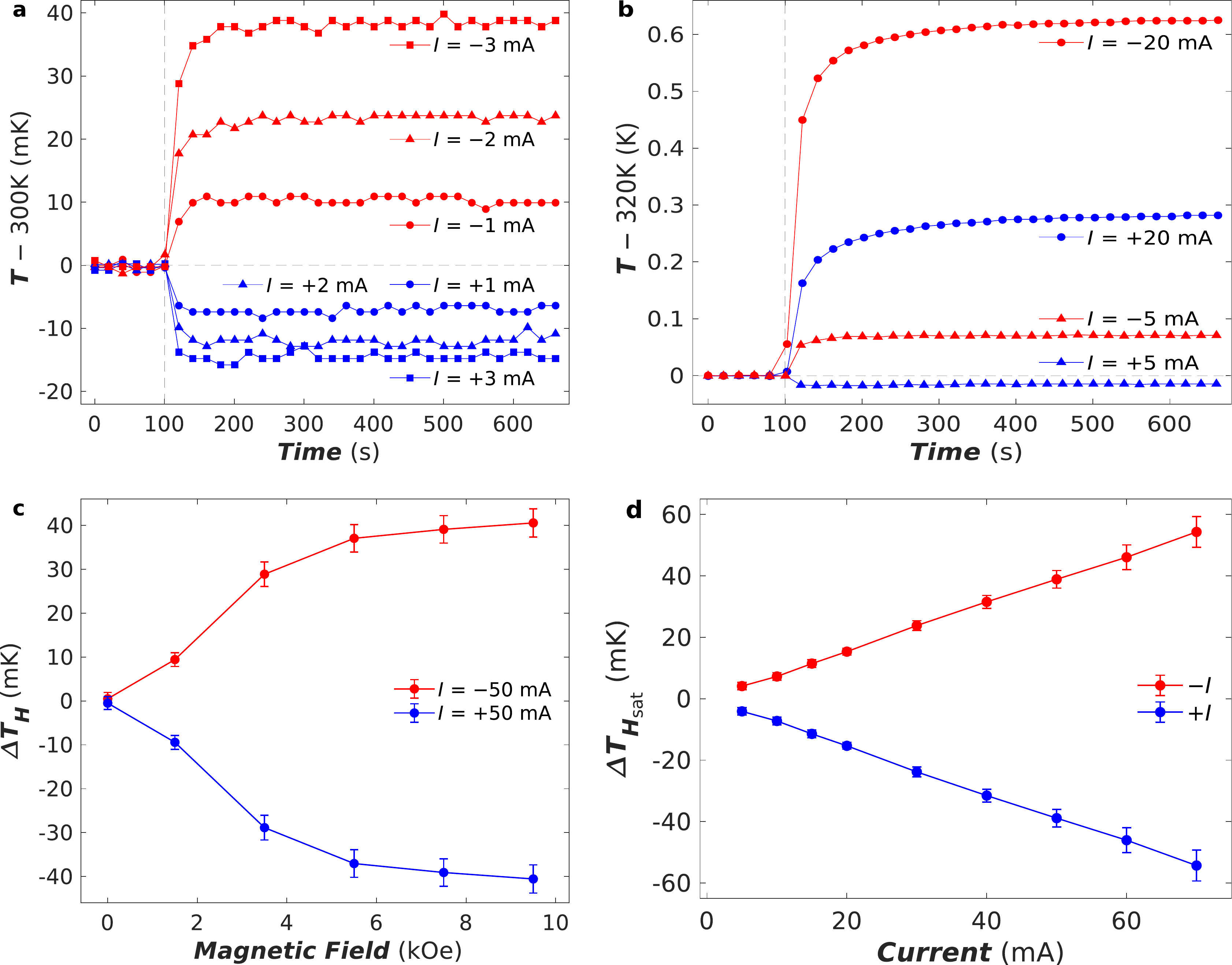}}{}
	\caption{{Direct observation of Peltier, Joule, current crowding, and magneto-Peltier effects. (a) Temperature vs. time traces of the sum of the Joule and Peltier heats relative to a working temperature of 300 K, as recorded by the Cernox sensor (see \fig\ref{Fig1}d). The direct currents (1-3 mA) are applied both forward and reverse in the interconnected Co/Cu nanowires of 105 nm in diameter and with a packing density of 22\% (R = 3.3 $\Omega$, MR = 12\%). (b) Same as in (a) but for higher current intensities for which the Peltier effect becomes dominated by the Joule heating. Both in (a) and (b), the DC current is switched on after 100 s as shown by the vertical dashed lines. (c) Measured temperature changes $\Delta T_H$ at the Peltier junction during the magnetic field sweep for a DC current of $-$50 mA and $+$50 mA. Here, the contribution from the Peltier heating has been estimated (Supporting Information Section S2). The Peltier term leads to heating and cooling at the saturation field of 9.5 kOe and depends on current flow direction. (d) Measured total temperature changes $\Delta T_{H_\text{sat}}$ at the Peltier junction between the zero-field ($T_{H_0}$) and saturated states ($T_{H_\text{sat}}$) vs current intensity applied both forward and reverse. The error bars in (c) and (d) reflect the uncertainty of the temperature measurements and is set to two times the standard deviation, gathering 95\% of the data variation.}}
	\label{Fig4}
\end{figure*}

Current flow in the high-packing density Co/Cu NW sample results in Joule heating as well as a Peltier heat current at the junctions between the NW network and the gold electrode, as shown in \fig\ref{Fig1}d. These heat flows were monitored continuously from the temperature changes using a Cernox sensor with respect to the operating temperatures for different current intensities and polarities, as well as for various applied magnetic fields. Net cooling occurs at low currents when the direction of the Peltier heat current flow is such that $\Pi I <$ 0 at the considered junction (with $\Pi$, the Peltier coefficient) and dominates over the Joule heating effect ($RI^2$), as shown in \fig\ref{Fig4}a. When the DC current flows from the NWs (with the higher Peltier coefficient) to the gold electrode (with the lower one), the Peltier heat is released from the junction, i.e. net cooling happens for $I \le$ $+$5 mA (see \fig\ref{Fig4}a and \ref{Fig4}b). As expected, the situation is reversed when the current flows in the opposite direction. For currents larger than 5 mA, the Joule heating dominates over the Peltier cooling, as shown in \fig\ref{Fig4}b by representative temperature vs. time traces for $I =$ $\pm$20 mA. Unlike the Peltier effect that leads to a cooling or a heating of the system depending on the direction of current flow, the Joule effect does not depend on the current polarity. It is thus possible to separate the two effects linearly (See Supporting Information Section S2). From \fig\ref{Fig4}b, one can estimate the Peltier cooling ability of the NW network of $\sim$8.5 K/A.

As shown in \fig\ref{Fig4}c, the magneto-Peltier effect has been quantified by recording the temperature change $\Delta T_H$ during the magnetic field sweeps. The field dependence of $\Delta T_H$ resembles to that of MR. The same measurements were performed for different currents, as shown in \fig\ref{Fig4}d where the total temperature change ($\Delta T_{H_\text{sat}}= T_{H_0}  - T_{H_\text{sat}} )$ between the zero-field ($T_{H_0}$) and saturation ($T_{H_\text{sat}}$) states is reported. As expected, the magneto-Peltier effect increases linearly with the driving current. From these results, one can obtain the guideline for the magnitude of the magnetically controlled cooling and heating ability of a macroscopic electronic component against the injected current. Using the values of $S_\uparrow$ and $S_\downarrow$ from \fig\ref{Fig3}b and the Onsager relation which relates the two thermoelectric coefficients $\Pi = ST$, one may estimate the maximum difference between the Peltier coefficients in the AP and P states for an infinite MR ratio as $\Pi_\uparrow - \Pi_\downarrow$, which correspond to -2.8 mV at RT for Co/Cu NWs. Since currents up to few hundreds mA are able to pass through the densely packed NW films without damaging the network structure, we may anticipate that a magnetic field can switch a heat flow as large as 1 mW.
	
\section{Conclusion}
In summary, we have fabricated large-area 3D interconnected Co/Cu multilayered nanowire networks exhibiting giant magnetoresistance and thermoelectric effects in the current-perpendicular-to-plane geometry. Our spin-caloritronic devices based on macroscopic nanowire networks overcome the insufficient power generation capability exhibited by the custom-patterned nanoscale magnetic structures reported previously. We have found very high thermoelectric power factor (PF) up to 7.5~mW/K$^2$m at room temperature, which are larger values than the PF of the widely used thermoelectric material, bismuth telluride. The PF can be magnetically modulated with PF change ratio of $\sim$100\% over a wide temperature range. Besides, large spin-dependent Seebeck and Peltier coefficients of -9.4~$\mu$V/K and -2.8 mV were obtained at room temperature, respectively. This capacity of net refrigerating effect controlled by a magnetic field holds promise for macroscopic light and flexible spin caloritronic devices. Our results should stimulate further studies for fabricating and designing practical thermo-electric coolers made of shapeable thermoelectric modules consisting of stacked nanowire network films that are connected electrically in series and thermally in parallel.
	
\section{Experimental Section} 
\noindent
\textit{\textbf{Sample fabrication}}: The polycarbonate (PC) porous membranes with interconnected pores have been fabricated by exposing a 22 $\mu$m thick PC film to a two-step irradiation process. The topology of the membranes is defined by exposing the film to a first irradiation step at two fixed angles of +25$^\circ$ and -25$^\circ$ with respect to the normal axis of the film plane. After rotating the PC film in the plane by 90$^\circ$, the second irradiation step takes place at the same fixed angular irradiation flux to finally form a 3D nanochannel network. The diameter of the latent tracks was enlarged by following a previously reported protocol to obtain membranes with distinct porosities and pores sizes \cite{Ferain2003}. The PC membranes with average pore diameter of 80 nm and 105 nm display low volumetric porosity (3\%) and large volumetric porosity (22\%), respectively. Next, the PC templates were coated on one side using an e-beam evaporator with a metallic Cr/Au bilayer to serve as cathode during the electrochemical deposition. The thickness of the thin adhesion layer of Cr was 3 nm, while for a uniform and consistent nanopore coverage withstanding the electrodeposition process, the Au film thickness was set to 400 nm and 750 nm for the 80 nm and 105 nm diameter porous membranes, respectively.
	
The multilayered NW networks have been grown at room temperature by electrodeposition into the 3D porous PC templates from a single sulphate bath using potentiostatic control and a pulsed electrodeposition technique \cite{Fert1999}. For these experiments, we used a Ag/AgCl reference electrode and a Pt counter electrode. To prepare the Co/Cu interconnected NW networks, the composition of the electrolyte was 1 M CoSO$_4$ $\cdot$ 7H$_2$O + 7.5 mM CuSO$_4$ $\cdot$ 5H$_2$O + 0.5 M H$_3$BO$_3$ and the deposition potential alternatively switched between $-0.95$ V to deposit Co layer containing approximately 5\% Cu impurity, and $-0.4$ V to deposit almost pure Cu layers \cite{Tang2007}. The nobler element (here Cu) is kept at much lower concentration than the magnetic element so that the rate of reduction of Cu is slow and limited by diffusion. As the electrodeposition of the different layers is performed in the same electrolyte, containing all the necessary cations, each individual layer of one type may contain some trace of the other. The potential is not high enough to deposit Co during the growth of  copper, but it is indeed to depose some copper with cobalt. In addition, the multilayered Co$_{0.95}$Cu$_{0.05}$/Cu nanowires were prepared using low pH sulfate bath in order to avoid magnetocrystalline contributions \cite{Darques2004}. Under these conditions, the magnetic properties are explained only from magnetostatic arguments like the shape anisotropy and dipolar coupling \cite{Medina2008_PhysRevB}.
Following a procedure described elsewhere \cite{Fert1999}, the deposition rates of each metals were determined from the pore filling time. According to this calibration, the deposition time was adjusted to 800 ms and 16 s for the Co and Cu layers, respectively, and the estimated average thickness of the bilayer was $\sim$15 nm, with approximately the same thicknesses for the Co and Cu layers. The morphology and elemental composition of the nano-structured interconnected NW networks were characterised using a high resolution field-emission scanning electron microscope (FE-SEM) JEOL 7600F equipped with an energy dispersive x-ray (EDX) analyser. The composition characterisation of the Co/Cu nanowire network by SEM-EDX leads an average Co content of 47\% and an average Cu content of 53\%. Since the Co layers contain about 5\% Cu impurity (as revealed by EDX analysis), the individual Co and Cu layer thicknesses are approximately the same. For the electron microscopy analysis, we removed the PC template by chemical dissolution using dichloromethane. For conducting magneto-transport measurements, the cathode was locally removed by plasma etching to create a two-probe design suitable for electric measurements, with the flow of current restricted along the nanowire segments, thus perpendicularly to the plane of the layers \cite{Camara-Santa-Clara-Gomes2016_JAP, Camara-Santa-Clara-Gomes2016_Nanoscale}.\\
	
\noindent
\textit{\textbf{Magneto-thermoelectric measurements}}: The thermoelectric power was measured by attaching one end of the sample to the copper sample holder using silver paint and a resistive heater to the other end. The voltage leads were made of thin Chromel P wires and the contribution of the leads to the measured thermoelectric power was subtracted out using the recommended values for the absolute thermopower of Chromel P. The temperature gradient was monitored with a small diameter type-E differential thermocouple. A typical temperature difference of 1 K was used in the measurements. The magneto-Seebeck measurements were made using low volumetric porosity PC membranes with nanowires partially filling the nanopores. A second set-up was built in order to unambiguously highlight Peltier effects in NW networks grown in high volumetric porosity PC membranes with the nanopores completely filled with nanowires to reduce the sample electrical resistance below 2 $\Omega$ and consequently the effects of Joule heating. In this case, a small Cernox thin film resistance sensor ($<3$~mg, 1 mm$^2$; Cernox-1010, Lake Shore Cryotronics Inc.) was attached at one junction between the NW network and the metal electrode. The temperature resolution of this highly sensitive thermometer is about 1 mK, which enables detection of Peltier-effect-based heating or cooling while a DC electrical current is flowing through the junction between the NW network and the metal electrode. All magneto-transport measurements were performed under vacuum with the magnetic field up to 1 T applied both in the plane and perpendicular to the plane of NW network films embedded in polymer membrane films. The temperature of the samples can be varied from 10~K to 320~K.
	
\section*{Acknowledgements} 
Financial support was provided by Wallonia/Brussels Community (ARC 13/18-052), and the Belgian Fund for Scientific Research (FNRS). F.A.A. is a Postdoctoral Researcher of the FNRS and T.daC.S.C.G is a Research Fellow of the FNRS. The authors would like to thank Dr. E. Ferain and the it4ip Company for supplying polycarbonate membranes.

\section{Supplementary Information}

\subsection{Estimation of the resistivity for Co/Cu NW network}

\begin{figure}[!h]
	\IfFileExists{FigS1.pdf}{\includegraphics[scale=0.5]{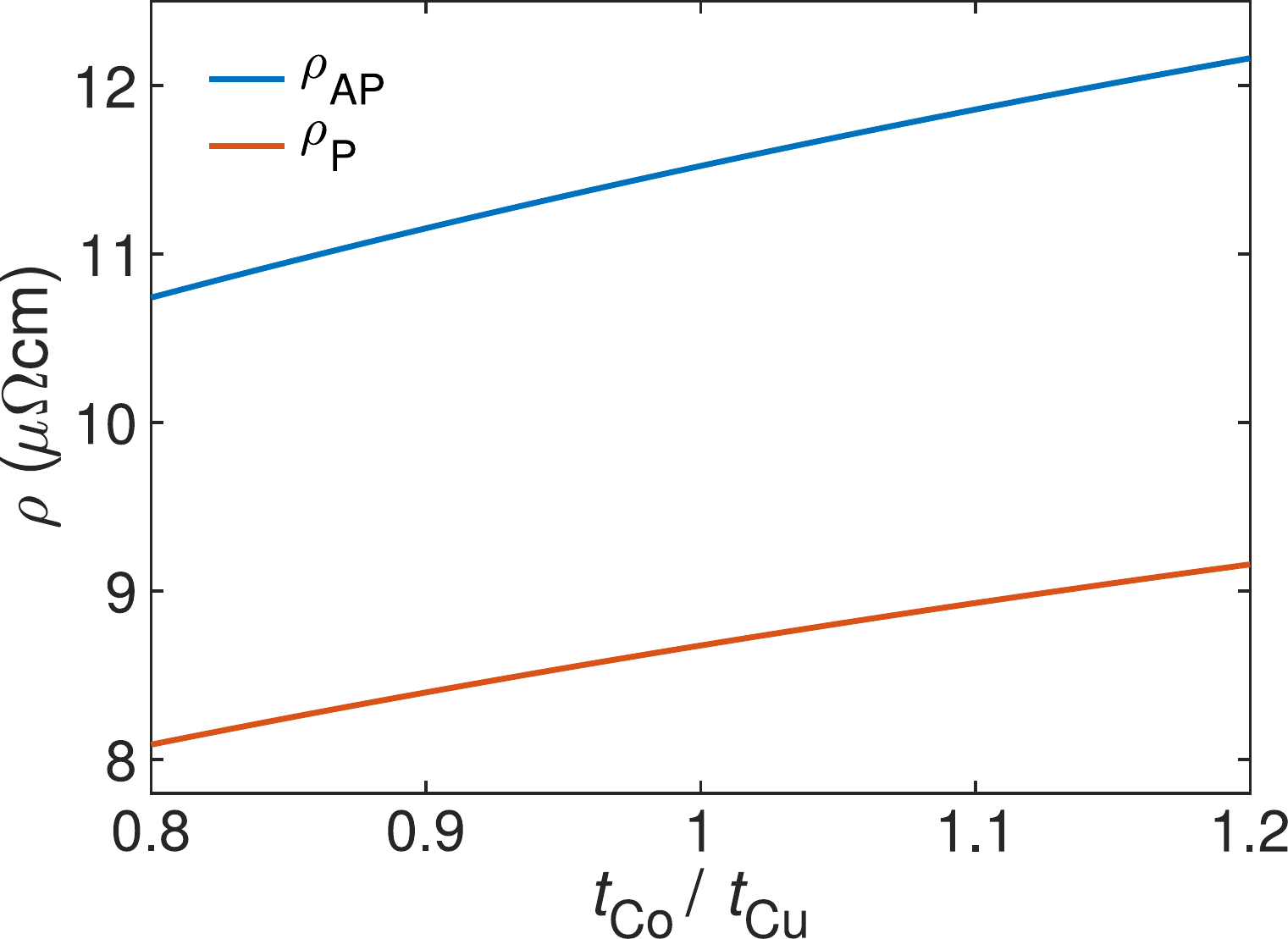}}{}
	\caption{Expected resistivity variations in the AP and P configurations as a function of the thickness ratio between Co and Cu layers.}
	\label{FigS1}
\end{figure}

Using the measured residual resistivity ratio $\text{RRR} = \rho_{\text{P}}(295 ~\text{K})/\rho_{\text{P}}(10 ~\text{K})$ for the 3D NW network samples in the parallel state, and considering a simple series resistance model for the multilayer as well as the Mathiessen's rule to separate the effects of thermally excited scatterings (phonons, magnons) and electron scattering by static defects (assuming that the interface resistance is independent of the temperature), the approximate expression for the room temperature (RT) resistivity is given by:
\begin{equation}
\rho_{\text{P}}^{\text{RT}} = \rho_{\text{P}}^0 + \frac{\rho_{\text{NM}}^{\text{RT}} t_{\text{NM}} + \rho_{\text{FM}}^{\text{RT}} t_{\text{FM}}}{t_{\text{NM}} + t_{\text{FM}}}\text{,}
\label{EqS1}
\end{equation}
where $t_{\text{NM}}$ and $t_{\text{FM}}$ are the thicknesses of the non-magnetic and magnetic layers respectively, $\rho_{\text{NM}}^{\text{RT}}$ and $\rho_{\text{FM}}^{\text{RT}}$ are the resistivities of the normal metal and ferromagnetic material at RT due thermally excited scatterings and $\rho_{\text{P}}^0$ the residual resistivity of the multilayer. The residual resistivity refers to interface resistance of Co/Cu multilayers and to scattering from impurities along with surface scattering within the NW network and internal grain-boundary scattering. Assuming that the resistivity values due to electron-phonon scattering of the nanowires closely correspond to bulk values (in a previous work on electrodeposited Ni nanowires\cite{Kamalakar2009}, it was shown that the electron-phonon coupling constant is nearly unchanged on size reduction down to 13 nm) and that thermally excited scatterings can be neglected at $T =$ 10 K (i.e., $\rho_P(10 ~ \text{K}) \approx \rho_{\text{P}}^0$), the above equation can be expressed in the form, using $t_{\text{NM}} \approx t_{\text{FM}}$:
\begin{equation}
\rho_{\text{P}}^{\text{RT}} \approx\frac{\rho_{\text{NM}}^{\text{RT}}  + \rho_{\text{FM}}^{\text{RT}} }{2 \big(1-\text{RRR}^{-1}\big)}\text{,}
\label{EqS2}
\end{equation}
so that it can also be easily obtained that:
\begin{equation}
\rho_{\text{AP}}^{\text{RT}} = \frac{\rho_{\text{P}}^{\text{RT}}}{1-\text{MR}}\text{.}
\label{EqS3}
\end{equation}
Using the RT resistivity values for the individual constituents from bulk material, i.e. $\rho_{\text{Cu}}^{\text{RT}} \approx 1.6$ $\mu \Omega$cm and $\rho_{\text{Co}}^{\text{RT}} \approx 6.2$ $\mu \Omega$cm and the measured RRR = 1.80, we estimated $\rho_{\text{AP}}^{\text{RT}} \approx$ 11.5 $\mu \Omega$cm for the interconnected network made of Co/Cu NWs 80 nm in diameter and 3\% packing density. This simple model based on the measured RRR ratio and intrinsic RT resistivity value is also consistent with previous results obtained on single nanowires\cite{Kamalakar2009, Karim2008}. In addition, any deviation in the strict equality in the layer thicknesses of the two constituent materials will only slightly change the estimated resistivity, as illustrated in \fig\ref{FigS1}.\\

\newcommand{\TJA}{\Delta{T}_{\text{J}}}%
\newcommand{\TJAerr}{\Delta{T}_{\text{J},\text{err}}}%

\newcommand{\TPA}{\Delta{T}_{\text{P}}}%
\newcommand{\TPAerr}{\Delta{T}_{\text{P},\text{err}}}%

\newcommand{\TsPzA}{\Delta{T}_{+}}%
\newcommand{\TsPzAerr}{\Delta{T}_{+,\text{err}}}%

\newcommand{\TsMzA}{\Delta{T}_{-}}%
\newcommand{\TsMzAerr}{\Delta{T}_{-,\text{err}}}%

\newcommand{\DMRA}{\text{MR}_H}%
\newcommand{\DMRAerr}{\text{MR}_{H,\text{err}}}%

\newcommand{\DMRAsat}{\text{MR}_{H_\text{sat}}}

\newcommand{\DPiRA}{\text{M}\Pi_H}%
\newcommand{\DPiRAerr}{\text{M}\Pi_{H,\text{err}}}%

\newcommand{\DPiRAsat}{\text{M}\Pi_{H_\text{sat}}}%

\newcommand{\DTsPA}{\Delta{T}_{+,H}}%
\newcommand{\DTsPAerr}{\Delta{T}_{+,H,\text{err}}}%

\newcommand{\DTsPAsat}{\Delta{T}_{+,H_\text{sat}}}%

\newcommand{\DTsMA}{\Delta{T}_{-,H}}%
\newcommand{\DTsMAerr}{\Delta{T}_{-,H,\text{err}}}%

\newcommand{\DTsMAsat}{\Delta{T}_{-,H_\text{sat}}}%

\subsection{Data on magneto-Peltier}

\begin{figure}[!htbp]
	\centering%
	\makeatletter%
	\IfFileExists{FigS2.pdf}{\includegraphics[scale=0.5]{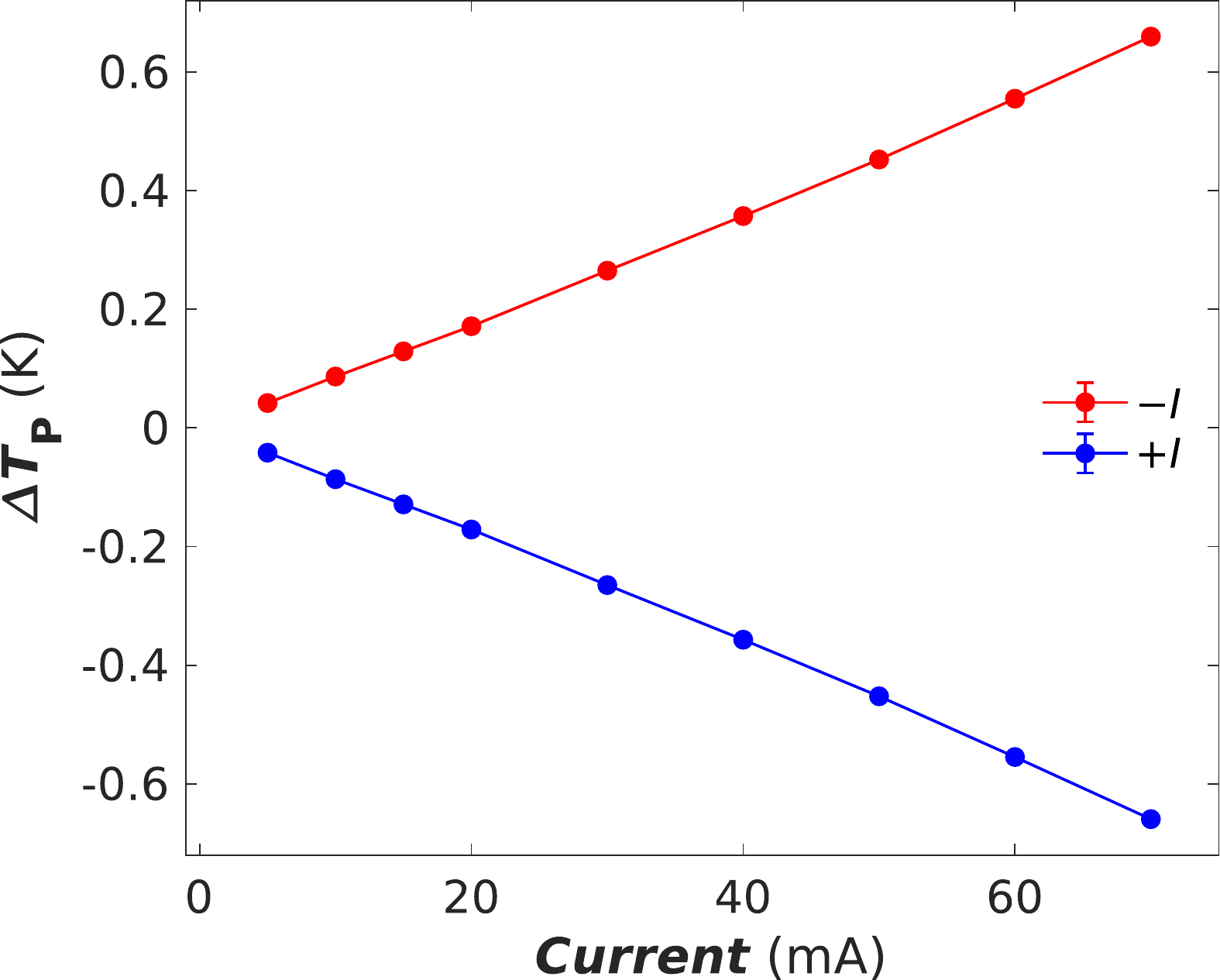}}{}
	\caption{Peltier temperature measurements. Measured temperature vs DC current intensity applied both forward and reverse as extracted using \eq{Eq:systB2} leading to the Peltier cooling or heating depending on the current polarity. The error bars reflect the uncertainty of the temperature measurements and is set to two times the standard deviation, gathering 95\% of the data variation.}
	\label{FigS2}
\end{figure}

\begin{figure*}[!hb]
	\makeatletter\IfFileExists{FigS3.pdf}{\includegraphics[scale=0.8]{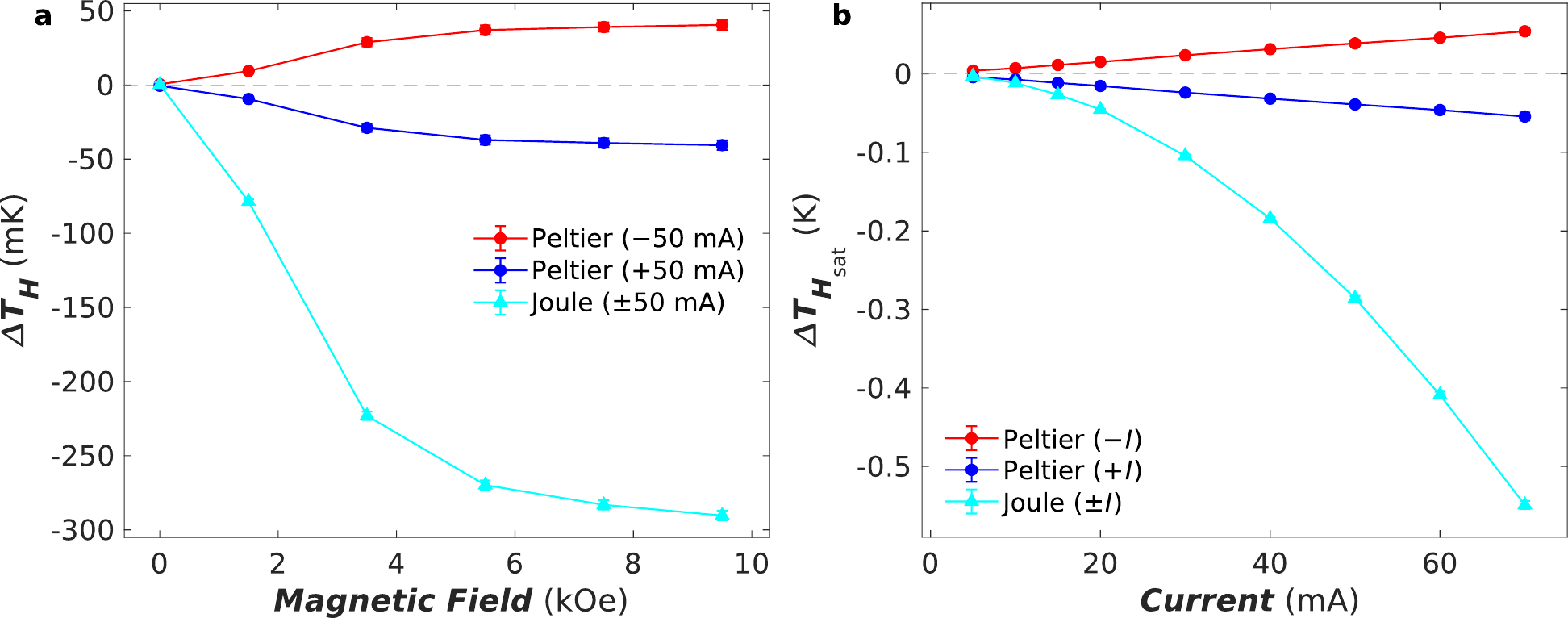}}{}
	\caption{Magneto-Joule and magneto-Peltier temperature measurements at the Peltier junction. (a) Measured temperature changes $\Delta T_H$ during the magnetic field sweep for a DC current of -50 mA and +50 mA. The Joule heating decrease due to the magneto-resistance effect is extracted combining \eqs{Eq:systB1} and \jeq{Eq:systB3}, and the Peltier term leads to heating or cooling, depending on the current polarity, as extracted combining \eqs{Eq:systB2} and \jeq{Eq:systB4}. (b) Measured total temperature changes $\Delta T_{H_\text{sat}}$ between the zero-field ($T_{H_0}$) and saturated states ($T_{H_\text{sat}}$) vs current intensity applied both forward and reverse as extracted combining \eqs{Eq:systB1} and \jeq{Eq:systB3sat} for the Joule contribution and combining \eqs{Eq:systB2} and \jeq{Eq:systB4sat} the Peltier contribution. The error bars reflect the uncertainty of the temperature measurements and is set to two times the standard deviation, gathering 95\% of the data variation.}
	\label{FigS3}
\end{figure*}

To separate the contributions of the Joule and the Peltier effects that take place simultaneously, two distinct measurements, one with forward and an other with reverse current polarity, are considered. The temperature change near the junction between the NW network and the gold electrode is extracted by measuring the resistance of the a Cernox thermometer integrated into the device, as depicted in Figure 2b, at a resolution of about 1 mK. In absence of an external magnetic field ($H_0$), the measured net temperature change relative to the working temperature ($T = 300$ K in Figure 4a; $T = 320$ K in \figs\ref{Fig4}(b-d), \ref{FigS2} and \ref{FigS3}), $\TsPzA$ and $\TsMzA$, corresponding to a positive and a negative DC current, respectively, can be written as:
\begin{gather}
\label{Eq:syst1}
\TJA+\TPA = \TsPzA,\\
\label{Eq:syst2}
\TJA-\TPA = \TsMzA,
\end{gather}
with $\TJA$ and $\TPA$, the temperature change due to the Joule and the Peltier effect, respectively. \eqs{Eq:syst1} and \jeq{Eq:syst2} simply lead to:
\begin{gather}
\label{Eq:systB1}
\TJA=\frac{\TsPzA+\TsMzA}{2},\\
\label{Eq:systB2}
\TPA=\frac{\TsPzA-\TsMzA}{2}.
\end{gather}

\fig\ref{FigS2} shows the Peltier temperature vs DC current from which one can extract the Peltier cooling/heating capacity of out sample, namely about 7.5 K/A, depending on the current polarity.\\

When a magnetic field $H$ is applied to the sample, the measured temperature changes $\DTsPA$ and $\DTsMA$ relative to the zero-field ($H_0$) measurements for a positive and a negative DC current, respectively, can be expressed as follows:
\begin{gather}
\label{Eq:syst3}
\DMRA\cdot\TJA+\DPiRA\cdot\TPA = \DTsPA,\\
\label{Eq:syst4}
\DMRA\cdot\TJA-\DPiRA\cdot\TPA = \DTsMA,
\end{gather}

with $\DMRA$ and $\DPiRA$, the magneto-resistance and the magneto-Peltier ratio at a given applied magnetic field $H$, respectively. Solving the linear systems of \eqs{Eq:syst3} and \jeq{Eq:syst4} yields:
\begin{gather}
\label{Eq:systB3}
\DMRA=\frac{\DTsPA+\DTsMA}{\TsPzA+\TsMzA},\\
\label{Eq:systB4}
\DPiRA=\frac{\DTsPA-\DTsMA}{\TsPzA-\TsMzA}.
\end{gather}

It should be noticed that at the saturation field $H_\text{sat}$, the magneto-resistance $\DMRAsat = (R_\text{AP} - R_\text{P})/R_\text{AP}$ and the magneto-Peltier $\DPiRAsat= (\Pi_\text{AP} - \Pi_\text{P})/\Pi_\text{AP}$ ratio can also be expressed as follows:
\begin{gather}
\label{Eq:systB3sat}
\DMRAsat = \frac{\DTsPAsat+\DTsMAsat}{\TsPzA+\TsMzA},\\
\label{Eq:systB4sat}
\DPiRAsat = \frac{\DTsPAsat-\DTsMAsat}{\TsPzA-\TsMzA}.
\end{gather}

Combining \eqs{Eq:systB2} and \jeq{Eq:systB4}, the product $\TPA\cdot\DPiRA$ leads to $\Delta T_H$ for different applied magnetic field values as plotted in Figure 4c and combining \eqs{Eq:systB2} and \jeq{Eq:systB4sat}, the product $\TPA\cdot\DPiRAsat$ leads to $\Delta T_{H_\text{sat}}$ at the saturation field $H_\text{sat}$ of 9.5 kOe for different values of the DC current as shown in Fig. 4D. The Joule counterparts originating from \eqs{Eq:systB1} and \jeq{Eq:systB3}, and \eqs{Eq:systB1} and \jeq{Eq:systB3sat} are shown in \figs\ref{FigS3}a and \ref{FigS3}b, respectively, and compared to the Peltier contribution. It can be observed that the Joules and Peltier contributions to $\Delta T_H$ have similar field dependence.\\

The results presented in \fig\ref{FigS3}b and \fig\ref{Fig4}d are consistent with the intuitive relation yielding characteristics of the form $\Delta T_{\pm} = RI^2 \pm \Pi I$, with the linear dependence vs the supplied current for the Peltier term and the quadratic behavior for the Joule term.

\bibliography{Refs}

\end{document}